\documentclass[twocolumn]{aastex62}

\newcommand{\lbol}{L_{\rm bol}}
\newcommand{\msun}{M_{\odot}}
\newcommand{\mbh}{M_{\rm BH}}
\newcommand{\nh}{N_{\rm H}}
\newcommand{\lx}{L_{2-10}}

\newcommand{\nustar}{\textit{NuSTAR}}

\usepackage{natbib}
\usepackage{amsmath}
\usepackage{multirow} %for using \multirow

\accepted{\today}
\shorttitle{Dead quasar engine in Arp 187}
\shortauthors{Ichikawa et al.}
\begin{document}

\title{\nustar\ Discovery of Dead Quasar Engine in Arp 187}

\correspondingauthor{Kohei Ichikawa and Taiki Kawamuro}
\email{k.ichikawa@astr.tohoku.ac.jp}
\email{taiki.kawamuro@nao.ac.jp}

\author[0000-0002-4377-903X]{Kohei Ichikawa}
\affil{Frontier Research Institute for Interdisciplinary Sciences, Tohoku University, Sendai 980-8578, Japan}
\affil{Astronomical Institute, Tohoku University, Aramaki, Aoba-ku, Sendai, Miyagi 980-8578, Japan}

\author[0000-0002-6808-2052]{Taiki Kawamuro}
\affil{National Astronomical Observatory of Japan, 2-21-1 Osawa, Mitaka, Tokyo 181-8588, Japan
}

\author{Megumi Shidatsu}
\affil{
Department of Physics, Faculty of Science, Ehime University, Matsuyama 790-8577, Japan
}

\author{Claudio Ricci}
\affiliation{
N\'ucleo de Astronom\'ia de la Facultad de Ingenier\'ia, Universidad Diego Portales, Av. Ej\'ercito Libertador 441, Santiago, Chile}
\affiliation{
Kavli Institute for Astronomy and Astrophysics, Peking University, Beijing 100871, China}

\author{Hyun-Jin Bae}
\affil{
Department of Medicine, University of Ulsan College of Medicine, Seoul 05505, Republic of Korea
}

\author{Kenta Matsuoka}
\affil{
Research Center for Space and Cosmic Evolution, Ehime University, 2-5 Bunkyo-cho, Matsuyama, Ehime 790-8577, Japan
}

\author{Jaejin Shin}
\affil{
Astronomy Program, Department of Physics and Astronomy, Seoul National University, Seoul, 151-742, Republic of Korea
}

\author{Yoshiki Toba}
\affil{
Department of Astronomy, Kyoto University, Kitashirakawa-Oiwake-cho, Sakyo-ku, Kyoto 606-8502, Japan
}
\affil{
Academia Sinica Institute of Astronomy and Astrophysics, 11F of Astronomy-Mathematics Building, AS/NTU, No.1, Section 4, Roosevelt Road, Taipei 10617, Taiwan
}
\affil{
Research Center for Space and Cosmic Evolution, Ehime University, 2-5 
Bunkyo-cho, Matsuyama, Ehime 790-8577, Japan
}

\author{Junko Ueda}
\affil{
National Astronomical Observatory of Japan, 2-21-1 Osawa, Mitaka, Tokyo 181-8588, Japan}
\affiliation{
Harvard-Smithsonian Center for Astrophysics, 60 Garden Street, Cambridge, MA 02138, USA}

\author{Yoshihiro Ueda}
\affil{
Department of Astronomy, Kyoto University, Kitashirakawa-Oiwake-cho, Sakyo-ku, Kyoto 606-8502, Japan
}



\begin{abstract}
Recent active galactic nucleus (AGN) and quasar surveys have revealed a population showing rapid AGN luminosity variability by a factor of $\sim10$. 
Here we present the most drastic AGN luminosity decline by a factor of $\gtrsim 10^{3}$ constrained by a $\nustar$ X-ray observation of the nearby galaxy Arp~187, which is a promising ``dead'' quasar whose current activity seems quiet but whose past activity of $\lbol \sim 
10^{46}$~erg~s$^{-1}$ is still observable at a large scale by its light echo.
The obtained upper bound of the X-ray luminosity is 
$\log (L_{\rm 2-10~keV}/{\rm erg}~{\rm s}^{-1}) < 41.2$,
corresponding to $\log (L_\mathrm{bol}/{\rm erg}~{\rm s}^{-1}) < 42.5$,
indicating an inactive central engine. 
Even if a putative torus model with $\nh \sim 1.5  \times 10^{24}$~cm$^{-2}$ is assumed, the strong upper-bound still holds with  $\log (L_{\rm 2-10~keV}/{\rm erg}~{\rm s}^{-1}) < 41.8$ or $\log (L_\mathrm{bol}/{\rm erg}~{\rm s}^{-1}) < 43.1$. Given the expected size of the narrow line region, this luminosity decrease by a factor of $\gtrsim 10^3$ must have occurred within $\lesssim 10^4$~yr.  
This extremely rapid luminosity/accretion shutdown is puzzling and it requires one burst-like accretion mechanism producing a clear outer boundary for an accretion disk.
We raise two possible scenarios realizing such an accretion mechanism: a mass accretion
1) by the tidal disruption of a molecular cloud 
and/or 
2) by the gas depletion as a result of vigorous nuclear 
starformation after rapid mass inflow to the central engine.

\end{abstract}

% currently to be inactive 
%disfavors the scenario of highly obscured AGN to 
%mitigate a big gap of the obtained luminosity 

\keywords{galaxies: active --- galaxies: nuclei ---
quasars: general}



\section{Introduction}\label{sec:intro}

%SMBH quench at some mass limit, why does it happen?
One of the fundamental questions on supermassive black holes (SMBHs) is how they stop growing their mass. 
The recent and ongoing quasar surveys have revealed
massive SMBHs with masses of $\mbh \gtrsim 10^9\ \msun$ at $z>7$ \citep[e.g.,][]{mor11}, 
and interestingly, there seems to be a redshift-independent maximum mass limit 
at $\mbh \sim 10^{10.5} \msun$ \citep[e.g.,][]{net03,kor13}.
%Theoretical prediction
This suggests that there is a fundamental
quenching mechanism of the SMBH growth independently from the cosmic evolution, and possible mechanisms have been discussed theoretically by several authors \citep[e.g.,][]{nat09,kin16,ina16}.

%Still, finding a dying AGN is still difficult.
However, it is still observationally difficult to find quasars in the final growing/dying phase. 
The Soltan argument requires the total AGN lifetime is 
the order of $10^{7-9}$ yr \citep{sol82,mar04}, and even a single episode of 
AGN activity should be longer than $10^5$~yr \citep{sch15}, and possibly $10^{6-7}$~yr \citep[e.g.,][]{mar04,hop06}.
This long lifetime implies that it is extremely difficult to witness the 
``newly-born'' or ``dying'' phase of each AGN within the human timescale of 
$\lesssim 100$~yr.

%Discovery/findings of fading AGN
One solution for this issue is using the difference in the physical size
among AGN indicators, some of which would give us the quasar time variability 
longer than the human timescale. AGN have multiple indicators with different 
physical scales from $10$--$100$~$R_{\rm g}$ \citep[X-ray corona and 
UV-optically bright accretion disk;][]{dai10,mor10}, $\sim 0.1$--$10$~pc
\citep[AGN tori;][]{bur13,ich15}, to $\sim 1-10$~kpc 
\citep[narrow-line region or AGN jet;][]{ben02,ode98}, and 
the luminosities of the AGN indicators are tightly correlated with each other
\citep{ich12,ich17a,ich19a,tob14,asm15,ued15}. 
Recent observations have revealed an interesting AGN population 
that shows strong AGN activity at large scales with $\sim$1 kpc but much weaker one at small scales ($<$ 10 pc), suggesting a fading activity of the central engine.
They are called fading AGN and currently $\sim20$ such sources have been reported \citep[e.g.,][]{sch13,ich16,ich19b,kee17,kaw17,vil18,wyl18,sar18b}.

%Introduction of Arp 187: one promissing dying AGN
Out of those $\sim20$ sources, Arp~187, a merger remnant infrared galaxy located
at $z=0.04$ ($D_L = 178$~Mpc), is the 
most promising ``dying'' or ``dead'' quasar candidate,
which completely lack current AGN signatures on small scales
($<10$~pc), but previous AGN activity estimated by the large scale AGN indicators ($\gtrsim 1$~kpc) must have reached quasar level luminosity.
Previous VLA and ALMA 5--100~GHz radio observations have revealed the bimodal jet lobes with $\sim5$~kpc size, whose kinematic jet age of $8\times10^4$~yr.
On the other hand, the central radio-core is absent, suggesting that the central engine
is already faint or even quenched.
The optical spectrum indicates that Arp~187 has
narrow line region with the estimated size of $\sim1$~kpc,
and the expected AGN luminosity reaches $L_\mathrm{bol}=
1.5 \times 10^{46}$~erg~s$^{-1}$  \citep{ich19b}.
On the nuclear AGN indicators, $\sim10$~pc scale AGN torus 
emission was not detected in the \textit{Spitzer}/IRS
mid-infrared spectrum, whose emission is dominated by the host galaxy, 
suggesting the absence of the current AGN
torus activity with the upper-bound of $L_\mathrm{bol} < 6 \times 10^{43}$~erg~s$^{-1}$ \citep{ich16}.

%This study: NuSTAR X-ray constraint
However, we still lack a strong constraint on the current activity.
In this letter, we report the first \nustar\ 
hard X-ray observation for this target. Thanks to its strong penetration power against absorption, \nustar\ puts the strongest constraint on the 
current AGN luminosity even in the case of heavy obscuration, allowing us to conclude that Arp~187 has an inactive central engine.

\section{\nustar\ Observations and Results}\label{sec:obs}

%------------------------------------------fig:~example-------------------------------------%
\begin{figure*}
\begin{center}
\includegraphics[width=1.0\textwidth]{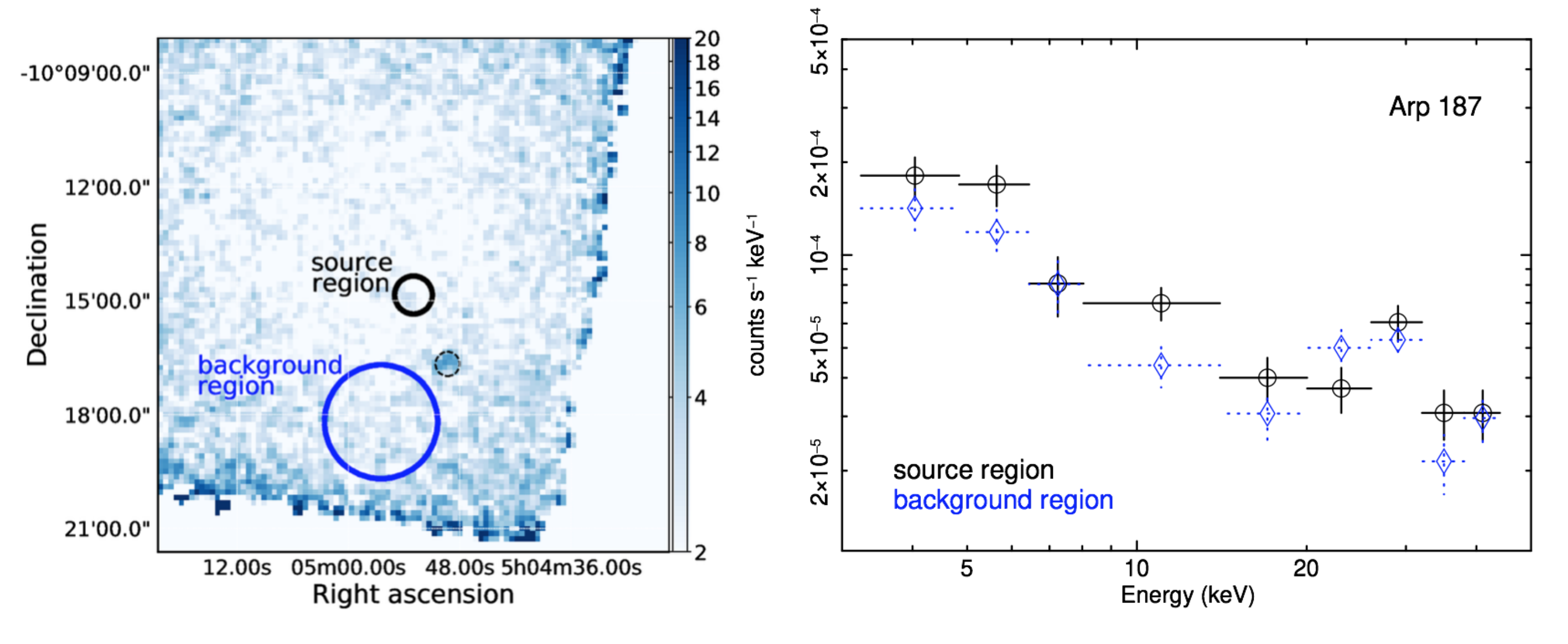}
\caption{
(Left) The exposure-corrected 8--24 keV image
of \nustar\ in units of 10$^{-6}$ counts s$^{-1}$ pixel$^{-1}$. This was created by combining the FPMA and FPMB data. 
The source/background region of Arp~187 is shown with black/blue solid circle, respectively.
The small dashed black circle represents an X-ray source, whose counterpart is likely to be GALEXASC J050449.00-101633.6, but is not our target Arp~187 (see text for more details).
(Right) The \nustar\ 3--50~keV spectra of the source and background regions 
(the black and blue circles in the left panel), indicated by the black
solid and blue dotted bins, respectively. 
}\label{fig:nustar}
\end{center}
\end{figure*}
%------------------------------------------fig:example-------------------------------------%

%NuSTAR data
The \nustar\ data were obtained with an on-source exposure of 82~ksec (GO cycle-4 
Program 04037, PI: K. Ichikawa). Following the ``\nustar~Analysis Quickstart Quide''
\footnote{\url{https://heasarc.gsfc.nasa.gov/docs/nustar/analysis/nustar_quickstart_guide.pdf}}, 
we reprocessed the data from \nustar~detector modules of FPMA and FPMB with the standard 
\nustar\ script of \texttt{nupipeline}, which has two options to remove times with high 
background (i.e., \texttt{saamode} and \texttt{tentable}). From the telemetry report on 
count rates over the focal plane, we found slightly higher rates in orbits around the 
standard SAA area ($\sim $ 2 counts s$^{-1}$) than typical values 
($\lesssim$ 1 count s$^{-1}$). 
Thus, \texttt{saamode=optimized} was adopted. 
Even if a more strict option of \texttt{saamode=strict} is used, our conclusion is unchanged. By contrast, such increase cannot be clearly seen in the so-called tentacle region \citep{for14} near the SAA, but by following recommendation of the \nustar\ team, we adopted \texttt{tentacle=yes}. Alternative option of \texttt{tentable=no} indeed provides a similar result, thus having little impact on our conclusion. The left panel of 
Figure~\ref{fig:nustar} shows an exposure-corrected 8--24 keV image, created by 
combining the FPMA and FPMB data and smoothed by a Gaussian function with $\sigma = 2$ pixels. 

%Background location
As indicated in the X-ray image, we defined a source region as a 
circle with a 30-arcsec radius centered at the optical position of the galaxy, and the 
background region was selected from the same chip as an off-source area with a 90~arcsec 
radius. The larger size was set to avoid local statistical fluctuations of the 
background level. We confirm insignificant change of our conclusion, even if a background 
spectrum is taken from a 30-arcsec circle near the source region. 
%A detection which is not associated with Arp 187
Note that, in the field-of-view, an X-ray source
was serendipitously detected in (R.A, Decl.)$\sim (05:04:49.325,-10:16:40.17)$ 
with $\approx$ 8.8$\sigma$ significance at 8--24 keV,
and its counterpart is likely to be
GALEXASC~J050449.00--101633.6
at $(05:04:49.0, -10:16:33.7)$. 
Its 2--10 keV flux estimated by a power-law model fit is $\sim 7\times10^{-14}$ 
erg cm$^{-2}$ s$^{-1}$.
Given its location far from our target Arp~187 with 
an angular separation of $\approx$ 2 arcmin,
which is at least six times larger than the positional uncertainty of \nustar~
\citep[up to $\simeq 20$~arcsec, e.g.,][]{lan17},
we conclude that the emission does not originate from Arp~187 and 
hereafter we will not discuss this source.

%NuSTAR X-ray spectra: Non-detection
The right panel of Figure~\ref{fig:nustar} shows obtained spectra
of Arp~187 at 3--50~keV 
from the two regions in the left panel. The source spectrum shows no significant 
excess (2.9$\sigma$ and 1.5$\sigma$ in the 3.0--8.0 keV and 8.0--24 keV bands, respectively) to the background one. By considering an un-absorbed cut-off power-law component with the photon index of 1.7 and cut-off energy of 360 keV 
\citep{kaw16b}\footnote{Even if we adopt another plausible parameter set of $\Gamma = 1.8$ and cut-off energy of 200 keV, found for a large hard X-ray selected AGN sample by \cite{ric17}, the upper limit of 2--10 keV luminosity increases only by $\approx$10\%, thus having little impact on our conclusion.}, the 3$\sigma$ 
upper limits of the 8--24 keV flux and luminosity are estimated to be 
$3.8\times10^{-14}$ erg~cm$^{-2}$~s$^{-1}$ and $1.4\times10^{41}$ erg~s$^{-1}$, equivalent
to the 2--10~keV luminosity of $1.6\times10^{41}$ erg~s$^{-1}$,
corresponding to $L_\mathrm{bol} < 3.2 \times 10^{42}$ erg~s$^{-1}$
 with a bolometric correction factor of 20 \citep[][]{Vas09}.
Hereafter, all upper-limits on X-ray fluxes are at $3\sigma$ level. 
 This estimate is not so sensitive to absorption in the sight-line up to $\log(N_{\rm H}/{\rm cm}^{-2}) \sim 23$. 
To consider more heavily obscured cases, we adopt a putative torus model as follows: 
%torusabs*zpowerlw*zhighect
\begin{verbatim} 
 TBabs*cabs*zpowerlw*zhighect
  +zpowerlw*zhighect
     *mtable{e-torus_20161121_2500M.fits} 
  +atable{refl_fe_torus.fits},  
\end{verbatim}
represented in XSPEC terminology\footnote{
The fits files of e-torus models were originally created by \cite{Ike09}. The first one is publicly available from  \url{https://heasarc.gsfc.nasa.gov/xanadu/xspec/models/etorus.html} and the second one was privately obtained from \cite{Ike09}}. 
This takes account of an absorbed and Compton scattered power-law component, a reflected continuum 
and an accompanying fluorescent iron-K$\alpha$ line.
The photon index of the power-law, inclination and opening angles of the torus are set to 1.7, 70$^\circ$, and 60$^\circ$, respectively. 
Even under a Compton-thick absorption of $\nh = 1.5 \times 10^{24}$~cm$^{-2}$ in the torus equatorial plane, the upper bound of the intrinsic luminosity is still very low with $\log (\lx/{\rm erg}~{\rm s}^{-1}) = 41.75$, or the bolometric luminosity of $\log 
(\lbol/{\rm erg}~{\rm s}^{-1}) = 43.05$. 
Note that other well-known torus models, 
such as MYTorus and Borus \citep[][]{Yaq12,Bal18},
also gives similar luminosity upper-bounds
with the difference by a factor of 1.2.
Lastly, we mention that the X-ray luminosity expected from the star-formation in the infrared \citep{ued14} is consistent with the 0.5--8 keV upper bound ($\sim 2\times10^{41}$ erg s$^{-1}$) from the extrapolation based on the 3--8 keV band, where a canonical power-law model seen in star-forming galaxies with $\Gamma=2.0$ and $3\times10^{21}$ cm$^{-2}$ \citep{min12} is utilized. 

\section{Discussion}\label{sec:discussion}

%------------------------------------------fig:~example-------------------------------------%
\begin{figure*}
\begin{center}
\includegraphics[width=0.80\textwidth]{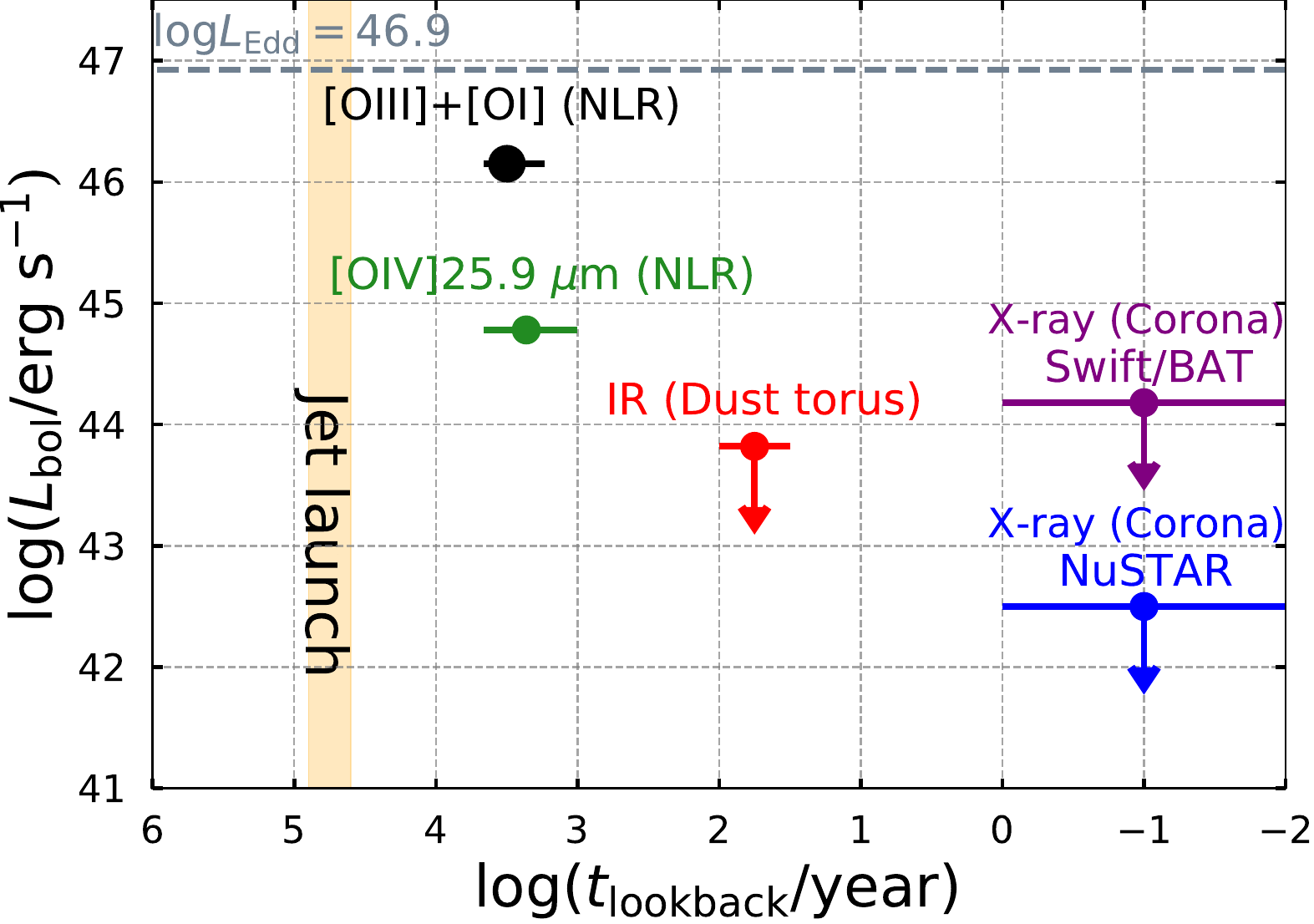}
\caption{
Long timescale light curve of Arp 187 based on the AGN indicators with multiple physical scales.
The estimated look back time is based on the light crossing time of each emission region except 
for ``jet launch'' time (orange area), which is estimated
to be $t_\mathrm{lookback}=8\times 10^{4}$~yr from the kinetic age of the jet lobe assuming 
its typical expansion speed of $v=0.1c$.
The all except blue point are taken from \cite{ich16,ich19b}.
The black/green point is obtained from the optical [\ion{O}{3}]$\lambda5007$+[\ion{O}{1}]$\lambda6300$
emission line and [\ion{O}{4}]25.89~$\mu$m emission.
The red point is obtained from the \textit{Spitzer}/IRS
spectra and the purple one is previously obtained
X-ray upper-bound from the \textit{Swift}/BAT hard X-ray
survey. The blue point shows the upper-bound
luminosity obtained by \nustar\ in this study.
The 3$\sigma$ upper-bounds are shown for the IR and X-ray observations.
}\label{fig:example}
\end{center}
\end{figure*}
%------------------------------------------fig:example-------------------------------------%

\subsection{Very Faint AGN Even If It Is Highly Obscured}

Our \nustar\ result shows the strongest current luminosity constraints with 
$\log (\lbol/\mathrm{erg~s}^{-1})< 42.5$ for $\log(\nh/{\rm cm}^{-2}) \lesssim 23$, and 
$\log (\lbol/\mathrm{erg~s}^{-1})< 43.1$ for $\log(\nh/{\rm cm}^{-2}) \simeq 24.2$. This indicates that the central engine of Arp~187 is currently very faint even if 
it is highly obscured by gas. This is consistent with the absence of the AGN 
torus emission in the \textit{Spitzer}/IRS spectra, 
which gives us the 3$\sigma$ upper-bound luminosity of 
$\log (\lbol/\mathrm{erg}~\mathrm{s}^{-1}) < 43.8$ \citep{ich16}.

One would expect that Arp~187 might be obscured by thicker absorption of $N_\mathrm{H}=10^{25}$~cm$^{-2}$. 
In this case, the expected upper-bound reaches to $\log (\lx/{\rm erg}~{\rm s}^{-1}) = 42.92$, or $\log (\lbol/{\rm erg}~{\rm s}^{-1}) = 44.22$, exceeding the upper-bound obtained from
the \textit{Spitzer}/IRS spectra.
However, this situation is unlikely because
the reprocessed infrared emission should be 
observed even in such highly obscured situation,
contributing to the \textit{Spitzer}/IRS spectra \citep[e.g.,][]{yan18}.
Thus, we conclude that the central engine of Arp~187
is likely to be dead, even if we consider the Compton-thick level obscuration, 
but the extreme absorption reaching $N_\mathrm{H}=10^{25}$~cm$^{-2}$ is also unlikely.

\subsection{The Drastic Luminosity Decline}

% Using the several AGN indicators, we
% can draw a lightcurve for over 10^4 yrs.

One important goal of our study is to
constrain how rapidly the AGN in Arp~187 has
dropped its luminosity.
As already described in~Section~\ref{sec:intro},
the mutli-wavelength observations indicate that
Arp~187 has experienced a luminosity decline in the past $\sim10^4$~yr.
Figure~\ref{fig:example} summarizes the long-term decline together 
with the X-ray upper-bound we have obtained. 
The luminosity and the look-back time are
obtained by combining the observational results of
several AGN indicators with different physical scales
\citep{ich16,ich19b}. 

Figure~\ref{fig:example} shows that, 
thanks to its sensitivity
in the hard X-ray band, \nustar\ (blue point) 
gives us a nearly two orders of magnitude fainter luminosity constraint than a previous estimate 
in the \textit{Swift}/BAT 105 month catalog 
\citep[purple; ][]{oh18}. 
In addition, the \nustar\ observation gives the constraint on the current luminosity better than the MIR observations. 
Compared to the luminosity of $\log (\lbol/{\rm erg}~{\rm s}^{-1})=46.15$ (see the black point) obtained from the NLR tracing the AGN activity $10^{3-4}$~yr ago, Arp~187 has experienced the  luminosity decline at least by a factor of $>10^3$.

%question: how can we reach such drastic luminosity decline?
Naively, this drastic luminosity experience indicates that the accretion rate in 
Arp~187 should have drastically dropped over $>10^3$ times within $10^4$~yr. 
This seemingly short timescale itself is consistent with the viscous timescale of the UV emitting region
\citep[see the discussion of][]{ich19b}. 
There however remains another question of how such drastic decline of accretion was  achieved. 
%intermittent accretion?
A gradual decrease of a external gas supply
to the accretion disk cannot produce such a drastic luminosity decline.
One suggestion is thus that the accretion disk has a 
clear outer disk boundary out of which the accretion rate drastically drops over $>10^3$ times.
Therefore, one burst-like accretion event is preferable for
realizing such a drastic accretion rate change.

\subsection{Tidal Disruption Event in Arp~187?}
%TDE of star? it is unlikely. 
One might argue that a tidal disruption event (TDE) of a star could reproduce such a drastic accretion change. 
However, there are three difficulties in the case of Arp~187. 
First, the estimated BH mass of Arp~187 is $6.7\times10^{8} 
\msun$, which thus requires a massive star above the main sequence, such as the red giant, to be tidally disrupted by the tidal field of the SMBH \citep[e.g.,][]{ree88}. 
The second is the luminosity problem: even if a red super giant, 
whose total mass is typically $\lesssim 50 \msun$, is tidally disrupted, 
it would be hard for the large BH ($\sim 7\times10^8 \msun$) 
to reach the expected Eddington ratio of Arp~187
($\lambda_{\rm Edd}\sim 0.1$), or an accretion rate 
of $\sim 2.5 \msun$/yr \citep[e.g., see Figure 5 of][]{mac13}.  
Third, the expected time scale: considering the rapid luminosity decline of TDEs which decays roughly as $L \propto t^{-5/3}$, the maximum observable timescale as AGN or quasar would be maximum $\lesssim 10$~yr. If a TDE is assumed to have happened at the time of jet-launch, or   $10^{4}$--$10^{5}$~yr ago (see Figure~\ref{fig:example} or Section~\ref{sec:intro}), the estimated NLR size should be expanded only up to $\sim 10$~pc scale, and the [\ion{O}{3}]  would cool on timescales of $\sim100$~yr and thus such feature is no longer observable at the current stage. 
This is in clear disagreement with the observations,
which leads us to exclude a TDE of a star as the
origin of the accretion episode currently observed in Arp~187.

% GMC cloud: possible?
The other possibility is the TDE of a giant molecular cloud (GMC). Arp 187 is a good environment to produce such an event because of the starforming galaxy with plenty gas mass of $\sim 2\times10^9 \msun$ in the central $\sim900$~pc \citep{ued14}.
The tidal radius of a GMC cloud is big enough as $R_{\rm TDE} = 200\times(R_{\rm GMC}/20~{\rm pc}) \times (M_{\rm BH}/10^8 M_{\odot})^{1/3} \times (M_{\rm GMC}/10^5 M_{\odot})^{-1/3}$ pc,
where a canonical range of GMC radii is $R_{\rm GMC} = 10$--$50$~pc and that of GMC masses is $M_{\rm GMC} = 10^4$--$10^6$~$M_{\rm sun}$ in local galaxies \citep[e.g.,][]{bol08}. Although this idea is exclusively
applied to Sgr A$\star$ \citep[e.g.,][]{bon08} and further theoretical studies are required to examine the case of much bigger SMBHs with $\mbh > 10^8 \msun$, a GMC with mass of $\sim10^6 \msun$ can
feed the SMBH of Arp~187 with the sub-Eddington level for $M_{\rm GMC}/(2.5 \msun~{\rm yr}^{-1}) \sim 4\times10^5$~yr. This would be long enough to produce the expected-size NLR by keeping the estimated past luminosity of $\log (L_{\rm bol}/{\rm erg}~{\rm s}^{-1}) \simeq 46.15$.

\subsection{Accretion Disk Outer Boundary After Nuclear Starburst}

Our observation indicates the rapid luminosity decline
in the final phase of quasar activity in Arp~187.
One question raised from this result is whether
this drastic luminosity decline is
unique event only for Arp~187 or 
a rather common behaviour in the final phase of quasars.

%Instead of or as a result of the falling of the GMC cloud(s),
Once the accretion rate somehow exceeds a certain value,
it may naturally
produce the drastic accretion rate gap,
resulting in the drastic luminosity decline in the final
phase of a quasar.
By utilizing the nuclear starburst disk model by \cite{tho05},
\cite{bal08} and \cite{ina16} discussed such a possibility
that once the rapid accretion rate of 
$>10 \msun$~yr$^{-1}$ is achieved, 
at around $\sim1-10$~pc,
vigorous star formation starts to deplete most of the
gas and the accretion rate rapidly decreases
by a factor of $\sim10^{2-3}$ times at some point, making a strong accretion rate gap. This is in good agreement with our expectation of the
clear outer accretion disk boundary. 

% The origin of such drastic rapid accretion rate
Considering that Arp~187 is a merger remnant,
such a rapid accretion flow with $>10 \msun$~yr$^{-1}$
could be achieved by a previous major merger
 \citep[e.g.,][]{hop10}.
% The expected AGN lifetime is also consistent with the previous observations.
The expected lifetime of such accretion disk is
$t_\mathrm{life} \sim t_\mathrm{vis} (r=1~\mathrm{pc}) \sim 5\times10^7$~yr, which is long enough to produce the NLR and actually consistent with the typical quasar lifetime
\citep[e.g.,][]{mar04b}. 
%Can we achieve such drastic accretion rate?
Based on those indirect observational signatures,
quasars who experienced a drastic accretion inflow
might follow the same luminosity decline in their future
after consuming most of the gas in the accretion disk.
On the other hand, a smooth accretion which have 
never exceeded the critical accretion rate of $\sim10 \msun$~yr$^{-1}$ will show more slower luminosity decline longer than $\sim10^4$~yr.

\acknowledgments

We acknowledge the anonymous referee for helpful suggestions that strengthened the paper.
We thank Mitsuru Kokubo, Ryo Tazaki, and Takuma Izumi for fruitful discussion.
This work is supported by Program for Establishing a Consortium
for the Development of Human Resources in Science
and Technology, Japan Science and Technology Agency (JST) and
 is partially supported by Japan Society for the Promotion of Science (JSPS) KAKENHI (18K13584; KI, 18J01050 and 19K14759; YT, 17K05384; YU). 
 T.K. was financially supported by the Grant-in-Aid for JSPS Fellows for young researchers (PD). 
 CR acknowledges support from the CONICYT+PAI Convocatoria Nacional subvencion a instalacion en la academia convocatoria a\~{n}o 2017 PAI77170080.


\vspace{5mm}
\facilities{\nustar}

\bibliographystyle{aasjournal}
\bibliography{arp187NuSTAR}

\begin{thebibliography}{}
\expandafter\ifx\csname natexlab\endcsname\relax\def\natexlab#1{#1}\fi
\providecommand{\url}[1]{\href{#1}{#1}}
\providecommand{\dodoi}[1]{doi:~\href{http://doi.org/#1}{\nolinkurl{#1}}}
\providecommand{\doeprint}[1]{\href{http://ascl.net/#1}{\nolinkurl{http://ascl.net/#1}}}
\providecommand{\doarXiv}[1]{\href{https://arxiv.org/abs/#1}{\nolinkurl{https://arxiv.org/abs/#1}}}

\bibitem[{{Asmus} {et~al.}(2015){Asmus}, {Gandhi}, {H{\"o}nig}, {Smette}, \&
  {Duschl}}]{asm15}
{Asmus}, D., {Gandhi}, P., {H{\"o}nig}, S.~F., {Smette}, A., \& {Duschl}, W.~J.
  2015, \mnras, 454, 766, \dodoi{10.1093/mnras/stv1950}

\bibitem[{{Ballantyne}(2008)}]{bal08}
{Ballantyne}, D.~R. 2008, \apj, 685, 787, \dodoi{10.1086/591048}

\bibitem[{{Balokovi{\'c}} {et~al.}(2018){Balokovi{\'c}}, {Brightman},
  {Harrison}, {Comastri}, {Ricci}, {Buchner}, {Gandhi}, {Farrah}, \&
  {Stern}}]{Bal18}
{Balokovi{\'c}}, M., {Brightman}, M., {Harrison}, F.~A., {et~al.} 2018, \apj,
  854, 42, \dodoi{10.3847/1538-4357/aaa7eb}

\bibitem[{{Bennert} {et~al.}(2002){Bennert}, {Falcke}, {Schulz}, {Wilson}, \&
  {Wills}}]{ben02}
{Bennert}, N., {Falcke}, H., {Schulz}, H., {Wilson}, A.~S., \& {Wills}, B.~J.
  2002, \apjl, 574, L105, \dodoi{10.1086/342420}

\bibitem[{{Bolatto} {et~al.}(2008){Bolatto}, {Leroy}, {Rosolowsky}, {Walter},
  \& {Blitz}}]{bol08}
{Bolatto}, A.~D., {Leroy}, A.~K., {Rosolowsky}, E., {Walter}, F., \& {Blitz},
  L. 2008, \apj, 686, 948, \dodoi{10.1086/591513}

\bibitem[{{Bonnell} \& {Rice}(2008)}]{bon08}
{Bonnell}, I.~A., \& {Rice}, W.~K.~M. 2008, Science, 321, 1060,
  \dodoi{10.1126/science.1160653}

\bibitem[{{Burtscher} {et~al.}(2013){Burtscher}, {Meisenheimer}, {Tristram},
  {Jaffe}, {H{\"o}nig}, {Davies}, {Kishimoto}, {Pott}, {R{\"o}ttgering},
  {Schartmann}, {Weigelt}, \& {Wolf}}]{bur13}
{Burtscher}, L., {Meisenheimer}, K., {Tristram}, K.~R.~W., {et~al.} 2013, \aap,
  558, A149, \dodoi{10.1051/0004-6361/201321890}

\bibitem[{{Dai} {et~al.}(2010){Dai}, {Kochanek}, {Chartas}, {Koz{\l}owski},
  {Morgan}, {Garmire}, \& {Agol}}]{dai10}
{Dai}, X., {Kochanek}, C.~S., {Chartas}, G., {et~al.} 2010, \apj, 709, 278,
  \dodoi{10.1088/0004-637X/709/1/278}

\bibitem[{{Forster} {et~al.}(2014){Forster}, {Harrison}, {Dodd}, {Stern},
  {Miyasaka}, {Madsen}, {Grefenstette}, {Markwardt}, {Craig}, \&
  {Marshall}}]{for14}
{Forster}, K., {Harrison}, F.~A., {Dodd}, S.~R., {et~al.} 2014, in Society of
  Photo-Optical Instrumentation Engineers (SPIE) Conference Series, Vol. 9149,
  \procspie, 91490R

\bibitem[{{Hopkins} {et~al.}(2006){Hopkins}, {Hernquist}, {Cox}, {Di Matteo},
  {Robertson}, \& {Springel}}]{hop06}
{Hopkins}, P.~F., {Hernquist}, L., {Cox}, T.~J., {et~al.} 2006, \apjs, 163, 1,
  \dodoi{10.1086/499298}

\bibitem[{{Hopkins} \& {Quataert}(2010)}]{hop10}
{Hopkins}, P.~F., \& {Quataert}, E. 2010, \mnras, 407, 1529,
  \dodoi{10.1111/j.1365-2966.2010.17064.x}

\bibitem[{{Ichikawa} {et~al.}(2017){Ichikawa}, {Ricci}, {Ueda}, {Matsuoka},
  {Toba}, {Kawamuro}, {Trakhtenbrot}, \& {Koss}}]{ich17a}
{Ichikawa}, K., {Ricci}, C., {Ueda}, Y., {et~al.} 2017, \apj, 835, 74,
  \dodoi{10.3847/1538-4357/835/1/74}

\bibitem[{{Ichikawa} {et~al.}(2019{\natexlab{a}}){Ichikawa}, {Ueda}, {Bae},
  {Kawamuro}, {Matsuoka}, {Toba}, \& {Shidatsu}}]{ich19b}
{Ichikawa}, K., {Ueda}, J., {Bae}, H.-J., {et~al.} 2019{\natexlab{a}}, \apj,
  870, 65, \dodoi{10.3847/1538-4357/aaf233}

\bibitem[{{Ichikawa} {et~al.}(2016){Ichikawa}, {Ueda}, {Shidatsu}, {Kawamuro},
  \& {Matsuoka}}]{ich16}
{Ichikawa}, K., {Ueda}, J., {Shidatsu}, M., {Kawamuro}, T., \& {Matsuoka}, K.
  2016, \pasj, 68, 9, \dodoi{10.1093/pasj/psv112}

\bibitem[{{Ichikawa} {et~al.}(2012){Ichikawa}, {Ueda}, {Terashima}, {Oyabu},
  {Gandhi}, {Matsuta}, \& {Nakagawa}}]{ich12}
{Ichikawa}, K., {Ueda}, Y., {Terashima}, Y., {et~al.} 2012, \apj, 754, 45,
  \dodoi{10.1088/0004-637X/754/1/45}

\bibitem[{{Ichikawa} {et~al.}(2015){Ichikawa}, {Packham}, {Ramos Almeida},
  {Asensio Ramos}, {Alonso-Herrero}, {Gonz{\'a}lez-Mart{\'{\i}}n},
  {Lopez-Rodriguez}, {Ueda}, {D{\'{\i}}az-Santos}, {Elitzur}, {H{\"o}nig},
  {Imanishi}, {Levenson}, {Mason}, {Perlman}, \& {Alsip}}]{ich15}
{Ichikawa}, K., {Packham}, C., {Ramos Almeida}, C., {et~al.} 2015, \apj, 803,
  57, \dodoi{10.1088/0004-637X/803/2/57}

\bibitem[{{Ichikawa} {et~al.}(2019{\natexlab{b}}){Ichikawa}, {Ricci}, {Ueda},
  {Bauer}, {Kawamuro}, {Koss}, {Oh}, {Rosario}, {Shimizu}, {Stalevski},
  {Fuller}, {Packham}, \& {Trakhtenbrot}}]{ich19a}
{Ichikawa}, K., {Ricci}, C., {Ueda}, Y., {et~al.} 2019{\natexlab{b}}, \apj,
  870, 31, \dodoi{10.3847/1538-4357/aaef8f}

\bibitem[{{Ikeda} {et~al.}(2009){Ikeda}, {Awaki}, \& {Terashima}}]{Ike09}
{Ikeda}, S., {Awaki}, H., \& {Terashima}, Y. 2009, \apj, 692, 608,
  \dodoi{10.1088/0004-637X/692/1/608}

\bibitem[{{Inayoshi} \& {Haiman}(2016)}]{ina16}
{Inayoshi}, K., \& {Haiman}, Z. 2016, \apj, 828, 110,
  \dodoi{10.3847/0004-637X/828/2/110}

\bibitem[{{Kawamuro} {et~al.}(2017){Kawamuro}, {Schirmer}, {Turner}, {Davies},
  \& {Ichikawa}}]{kaw17}
{Kawamuro}, T., {Schirmer}, M., {Turner}, J.~E.~H., {Davies}, R.~L., \&
  {Ichikawa}, K. 2017, \apj, 848, 42, \dodoi{10.3847/1538-4357/aa8e46}

\bibitem[{{Kawamuro} {et~al.}(2016){Kawamuro}, {Ueda}, {Tazaki}, {Ricci}, \&
  {Terashima}}]{kaw16b}
{Kawamuro}, T., {Ueda}, Y., {Tazaki}, F., {Ricci}, C., \& {Terashima}, Y. 2016,
  \apjs, 225, 14, \dodoi{10.3847/0067-0049/225/1/14}

\bibitem[{{Keel} {et~al.}(2017){Keel}, {Lintott}, {Maksym}, {Bennert},
  {Chojnowski}, {Moiseev}, {Smirnova}, {Schawinski}, {Sartori}, {Urry},
  {Pancoast}, {Schirmer}, {Scott}, {Showley}, \& {Flatland}}]{kee17}
{Keel}, W.~C., {Lintott}, C.~J., {Maksym}, W.~P., {et~al.} 2017, \apj, 835,
  256, \dodoi{10.3847/1538-4357/835/2/256}

\bibitem[{{King}(2016)}]{kin16}
{King}, A. 2016, \mnras, 456, L109, \dodoi{10.1093/mnrasl/slv186}

\bibitem[{{Kormendy} \& {Ho}(2013)}]{kor13}
{Kormendy}, J., \& {Ho}, L.~C. 2013, \araa, 51, 511,
  \dodoi{10.1146/annurev-astro-082708-101811}

\bibitem[{{Lansbury} {et~al.}(2017){Lansbury}, {Stern}, {Aird}, {Alexander},
  {Fuentes}, {Harrison}, {Treister}, {Bauer}, {Tomsick}, {Balokovi{\'c}}, {Del
  Moro}, {Gandhi}, {Ajello}, {Annuar}, {Ballantyne}, {Boggs}, {Brandt},
  {Brightman}, {Chen}, {Christensen}, {Civano}, {Comastri}, {Craig}, {Forster},
  {Grefenstette}, {Hailey}, {Hickox}, {Jiang}, {Jun}, {Koss}, {Marchesi},
  {Melo}, {Mullaney}, {Noirot}, {Schulze}, {Walton}, {Zappacosta}, \&
  {Zhang}}]{lan17}
{Lansbury}, G.~B., {Stern}, D., {Aird}, J., {et~al.} 2017, \apj, 836, 99,
  \dodoi{10.3847/1538-4357/836/1/99}

\bibitem[{{MacLeod} {et~al.}(2013){MacLeod}, {Ramirez-Ruiz}, {Grady}, \&
  {Guillochon}}]{mac13}
{MacLeod}, M., {Ramirez-Ruiz}, E., {Grady}, S., \& {Guillochon}, J. 2013, \apj,
  777, 133, \dodoi{10.1088/0004-637X/777/2/133}

\bibitem[{{Marconi} {et~al.}(2004){Marconi}, {Risaliti}, {Gilli}, {Hunt},
  {Maiolino}, \& {Salvati}}]{mar04}
{Marconi}, A., {Risaliti}, G., {Gilli}, R., {et~al.} 2004, \mnras, 351, 169,
  \dodoi{10.1111/j.1365-2966.2004.07765.x}

\bibitem[{{Martini}(2004)}]{mar04b}
{Martini}, P. 2004, in Coevolution of Black Holes and Galaxies, ed. L.~C. {Ho},
  169

\bibitem[{{Mineo} {et~al.}(2012){Mineo}, {Gilfanov}, \& {Sunyaev}}]{min12}
{Mineo}, S., {Gilfanov}, M., \& {Sunyaev}, R. 2012, \mnras, 419, 2095,
  \dodoi{10.1111/j.1365-2966.2011.19862.x}

\bibitem[{{Morgan} {et~al.}(2010){Morgan}, {Kochanek}, {Morgan}, \&
  {Falco}}]{mor10}
{Morgan}, C.~W., {Kochanek}, C.~S., {Morgan}, N.~D., \& {Falco}, E.~E. 2010,
  \apj, 712, 1129, \dodoi{10.1088/0004-637X/712/2/1129}

\bibitem[{{Mortlock} {et~al.}(2011){Mortlock}, {Warren}, {Venemans}, {Patel},
  {Hewett}, {McMahon}, {Simpson}, {Theuns}, {Gonz{\'a}les-Solares}, {Adamson},
  {Dye}, {Hambly}, {Hirst}, {Irwin}, {Kuiper}, {Lawrence}, \&
  {R{\"o}ttgering}}]{mor11}
{Mortlock}, D.~J., {Warren}, S.~J., {Venemans}, B.~P., {et~al.} 2011, \nat,
  474, 616, \dodoi{10.1038/nature10159}

\bibitem[{{Natarajan} \& {Treister}(2009)}]{nat09}
{Natarajan}, P., \& {Treister}, E. 2009, \mnras, 393, 838,
  \dodoi{10.1111/j.1365-2966.2008.13864.x}

\bibitem[{{Netzer}(2003)}]{net03}
{Netzer}, H. 2003, \apjl, 583, L5, \dodoi{10.1086/368012}

\bibitem[{{O'Dea}(1998)}]{ode98}
{O'Dea}, C.~P. 1998, \pasp, 110, 493, \dodoi{10.1086/316162}

\bibitem[{{Oh} {et~al.}(2018){Oh}, {Koss}, {Markwardt}, {Schawinski},
  {Baumgartner}, {Barthelmy}, {Cenko}, {Gehrels}, {Mushotzky}, {Petulante},
  {Ricci}, {Lien}, \& {Trakhtenbrot}}]{oh18}
{Oh}, K., {Koss}, M., {Markwardt}, C.~B., {et~al.} 2018, ArXiv e-prints.
\newblock \doarXiv{1801.01882}

\bibitem[{{Rees}(1988)}]{ree88}
{Rees}, M.~J. 1988, \nat, 333, 523, \dodoi{10.1038/333523a0}

\bibitem[{{Ricci} {et~al.}(2017){Ricci}, {Trakhtenbrot}, {Koss}, {Ueda}, {Del
  Vecchio}, {Treister}, {Schawinski}, {Paltani}, {Oh}, {Lamperti}, {Berney},
  {Gandhi}, {Ichikawa}, {Bauer}, {Ho}, {Asmus}, {Beckmann}, {Soldi},
  {Balokovi{\'c}}, {Gehrels}, \& {Markwardt}}]{ric17}
{Ricci}, C., {Trakhtenbrot}, B., {Koss}, M.~J., {et~al.} 2017, \apjs, 233, 17,
  \dodoi{10.3847/1538-4365/aa96ad}

\bibitem[{{Sartori} {et~al.}(2018){Sartori}, {Schawinski}, {Trakhtenbrot},
  {Caplar}, {Treister}, {Koss}, {Urry}, \& {Zhang}}]{sar18b}
{Sartori}, L.~F., {Schawinski}, K., {Trakhtenbrot}, B., {et~al.} 2018, \mnras,
  476, L34, \dodoi{10.1093/mnrasl/sly025}

\bibitem[{{Schawinski} {et~al.}(2015){Schawinski}, {Koss}, {Berney}, \&
  {Sartori}}]{sch15}
{Schawinski}, K., {Koss}, M., {Berney}, S., \& {Sartori}, L.~F. 2015, \mnras,
  451, 2517, \dodoi{10.1093/mnras/stv1136}

\bibitem[{{Schirmer} {et~al.}(2013){Schirmer}, {Diaz}, {Holhjem}, {Levenson},
  \& {Winge}}]{sch13}
{Schirmer}, M., {Diaz}, R., {Holhjem}, K., {Levenson}, N.~A., \& {Winge}, C.
  2013, \apj, 763, 60, \dodoi{10.1088/0004-637X/763/1/60}

\bibitem[{{Soltan}(1982)}]{sol82}
{Soltan}, A. 1982, \mnras, 200, 115, \dodoi{10.1093/mnras/200.1.115}

\bibitem[{{Thompson} {et~al.}(2005){Thompson}, {Quataert}, \& {Murray}}]{tho05}
{Thompson}, T.~A., {Quataert}, E., \& {Murray}, N. 2005, \apj, 630, 167,
  \dodoi{10.1086/431923}

\bibitem[{{Toba} {et~al.}(2014){Toba}, {Oyabu}, {Matsuhara}, {Malkan},
  {Gandhi}, {Nakagawa}, {Isobe}, {Shirahata}, {Oi}, {Ohyama}, {Takita},
  {Yamauchi}, \& {Yano}}]{tob14}
{Toba}, Y., {Oyabu}, S., {Matsuhara}, H., {et~al.} 2014, \apj, 788, 45,
  \dodoi{10.1088/0004-637X/788/1/45}

\bibitem[{{Ueda} {et~al.}(2014){Ueda}, {Iono}, {Yun}, {Crocker}, {Narayanan},
  {Komugi}, {Espada}, {Hatsukade}, {Kaneko}, {Matsuda}, {Tamura}, {Wilner},
  {Kawabe}, \& {Pan}}]{ued14}
{Ueda}, J., {Iono}, D., {Yun}, M.~S., {et~al.} 2014, \apjs, 214, 1,
  \dodoi{10.1088/0067-0049/214/1/1}

\bibitem[{{Ueda} {et~al.}(2015){Ueda}, {Hashimoto}, {Ichikawa}, {Ishino},
  {Kniazev}, {V{\"a}is{\"a}nen}, {Ricci}, {Berney}, {Gandhi}, {Koss},
  {Mushotzky}, {Terashima}, {Trakhtenbrot}, \& {Crenshaw}}]{ued15}
{Ueda}, Y., {Hashimoto}, Y., {Ichikawa}, K., {et~al.} 2015, \apj, 815, 1,
  \dodoi{10.1088/0004-637X/815/1/1}

\bibitem[{{Vasudevan} {et~al.}(2009){Vasudevan}, {Mushotzky}, {Winter}, \&
  {Fabian}}]{Vas09}
{Vasudevan}, R.~V., {Mushotzky}, R.~F., {Winter}, L.~M., \& {Fabian}, A.~C.
  2009, \mnras, 399, 1553, \dodoi{10.1111/j.1365-2966.2009.15371.x}

\bibitem[{{Villar-Mart{\'{\i}}n} {et~al.}(2018){Villar-Mart{\'{\i}}n},
  {Cabrera-Lavers}, {Humphrey}, {Silva}, {Ramos Almeida}, {Piqueras-L{\'o}pez},
  \& {Emonts}}]{vil18}
{Villar-Mart{\'{\i}}n}, M., {Cabrera-Lavers}, A., {Humphrey}, A., {et~al.}
  2018, \mnras, 474, 2302, \dodoi{10.1093/mnras/stx2911}

\bibitem[{{Wylezalek} {et~al.}(2018){Wylezalek}, {Zakamska}, {Greene},
  {Riffel}, {Drory}, {Andrews}, {Merloni}, \& {Thomas}}]{wyl18}
{Wylezalek}, D., {Zakamska}, N.~L., {Greene}, J.~E., {et~al.} 2018, \mnras,
  474, 1499, \dodoi{10.1093/mnras/stx2784}

\bibitem[{{Yan} {et~al.}(2019){Yan}, {Hickox}, {Hainline}, {Stern}, {Lansbury},
  {Alexander}, {Hviding}, {Assef}, {Ballantyne}, {Dipompeo}, {Lanz}, {Carroll},
  {Koss}, {Lamperti}, {Civano}, {Del Moro}, {Gandhi}, \& {Myers}}]{yan18}
{Yan}, W., {Hickox}, R.~C., {Hainline}, K.~N., {et~al.} 2019, \apj, 870, 33,
  \dodoi{10.3847/1538-4357/aaeed4}

\bibitem[{{Yaqoob}(2012)}]{Yaq12}
{Yaqoob}, T. 2012, \mnras, 423, 3360, \dodoi{10.1111/j.1365-2966.2012.21129.x}

\end{thebibliography}
%\bibliography{Arp187NuSTARtest}
%\end{thebibliography}



\end{document}